\documentclass[proof]{WileyASNA-v1}

\articletype{Article Type}%

\received{28 March 2025}
\revised{28 March 2025}
\accepted{28 March 2025}

\begin{document}


\title{Stellar Evolution in Close Binaries: Processes and Outcomes}

\author[1,2]{Omar Gustavo Benvenuto}
\author[1,2]{María Alejandra De Vito}
\author[1,2]{Leandro Bartolomeo Koninckx}
\author[1,3]{Maite Echeveste}
\author[1,2]{María Leonela Novarino}
\author[4]{Jorge Ernesto Horvath}

\authormark{O. G. Benvenuto \textsc{et al.}}

\address[1]{\orgdiv{Instituto de Astrofísica de La Plata}, \orgname{CCT CONICET-UNLP}, \orgaddress{\state{Paseo del Bosque S/N, 1900 La Plata}, \country{Argentina}}}

\address[2]{\orgdiv{Facultad de Ciencias Astronómicas y Geofísicas}, \orgname{Universidad Nacional de La Plata}, \orgaddress{\state{Paseo del Bosque S/N, 1900 La Plata}, \country{Argentina}}}

\address[3]{\orgdiv{Osservatorio Astrofisico di Arcetri}, \orgname{INAF}, \orgaddress{\state{Largo Enrico Fermi 5,  I-50125 Firenze}, \country{Italy}}}

\address[4]{\orgdiv{Instituto de Astronomia, Geof\'\i sica e Ci\^encias Atmosf\'ericas}, \orgname{IAG-USP}, \orgaddress{\state{Rua do Mat\~ao 1226, S\~ao Paulo SP}, \country{Brazil}}}

\corres{*Corresponding author name. \email{obenvenu@fcaglp.unlp.edu.ar}}


\abstract{We discuss some aspects of stellar evolution in binary systems. While single stars can swell following the chemical evolution of their interior, stars belonging to binary systems cannot overflow the size of the Roche lobe and hydrostatic equilibrium is strictly impossible. The system is forced to exchange mass between its members through the inner Lagrangian point. In the first part of the paper, we discuss the standard evolution of binaries that have a non-degenerate donor star and a compact companion. We show that the model fails when to account for the occurrence of binary pulsars when they predict a long-standing mass transfer episode. Models including irradiation feedback and evaporation in close binaries are examined next. Following these sections, we discuss the case of systems with a black hole (BH). We show that if BHs are born non-rotating, binary interaction seems insufficient to speed them up, an indication that BH's rotation is a feature present at birth. Finally, we discuss Blue Straggler Stars detected in open and globular clusters. Since they cannot be understood as single-born stars, we evaluate one of the proposed channels is mass transfer in close binaries, and discuss its viability and the limitations of the present models.}


\maketitle

\footnotetext{\textbf{Abbreviations:} BH, black hole; BSS, blue straggler star; BW, black widow pulsar; CBS, close binary system; {\tt GESBI}, Group of Study of Interacting Binary Systems; HMXB (LMXB), high (low) mass X-ray binary; HRD, Hertzsprung-Russell diagram; IFB, irradiation feedback; NS, neutron star; RB, redback pulsar.}

\section{Introduction}\label{sec:intro}

Since the beginning of the past century, it became clear that stars are stable nuclear fusion cauldrons in hydrostatic equilibrium that play a central role in the production of most elements in Nature \citep{Clayton}. This is one of the most remarkable achievements of Science. Stellar evolution is a direct consequence of the joint action of several physical ingredients, e.g., nuclear reaction rates, radiative and conductive opacities, an equation of state (including partial ionizations, electron degeneracy, Coulomb interactions, etc.), non-adiabatic convection, and wind mass loss, among others. Although the present understanding of these ingredients is still far from complete, it allows us to state that the evolution of single stars is well understood. For a description of these topics, see e.g., \citet{1990sse..book.....K}. Despite this fact, the modeling of single-star evolution is still not fully predictive. Among the uncertainties that affect the models, perhaps the most serious one is related to the still poor treatment of convection. But it is important to stress that the enormous development of Physics rendered as a {\it bonus} and a consistent picture of Stellar Evolution as a prime application field. When the more complex context of binary systems is considered, not surprisingly, Stellar Evolution is not so well understood. The interplay of physical ingredients has to be disentangled, and accurate observations play a major role in modeling and understanding different classes of binaries.

This paper aims to review the research work carried out in our working group {\tt GESBI} and some general issues. Here, we describe some of the main ingredients that determine Stellar Evolution in binary systems and related astrophysical objects.  Firstly, in Section~\ref{Sec:binarias}, we describe the usual approximations to make binary evolution tractable with simple 1D simulations. Then, we present the case in which one of the components of the system is a neutron star (NS). In Section~\ref{Sec:conNSs}, we describe the usual approximations employed to treat these systems that are responsible for the occurrence of binary millisecond pulsars and X-ray sources. Since these models have severe difficulties in accounting for the existence of some objects, it is necessary to include non-standard ingredients, such as irradiation and evaporation (which are strongly suggested by observations). This is addressed in Section~\ref{sec:Ifeedback}. Section~\ref{Sec:conBHs} is devoted to the discussion of the case in which the compact companion is a black hole (BH). The case in which both stars are normal\footnote{We refer to ``normal'' stars those composed of non-degenerate gas, i.e., not compact objects.}, the low-mass star case is presented in Section~\ref{sec:BSSs} in connection with the existence of blue straggler stars (BSSs). Lack of space prevents us from discussing other astrophysical systems related to binary evolution, e.g., cataclysmic variables (in which one of the stars is a white dwarf), and progenitors of supernovae.

\section{The general context of binary evolution}\label{Sec:binarias}

It is usual to study binary evolution considering the circular, restricted three-body problem, assuming that stellar masses are concentrated in points. This is a very good approximation in this context; see, e.g., \citet{puntos_lagrangianos}. The shape of the stars has to be that of the equipotential surfaces. As it is well-known, these surfaces can be described analytically. In a corotating reference frame, there are five {\it Lagrangian} points at which the net force on any test particle is zero. Particularly relevant is the one located in between the stars, usually called $L_{1}$. The equipotential surfaces that surround each star and are connected at $L_{1}$ define the {\it Roche Lobes}.

Binary configurations are generally classified into different types \citep{puntos_lagrangianos}. Binaries may be {\it detached} if both stars accommodate inside their respective Roche lobes. If only one of them fills its lobe, this is called {\it semi-detached}. If both stars fill their lobes, these are referred to as {\it contact} binaries. While isolated stars may swell freely, in binary systems there is a limit imposed by the Roche lobes.

The size of the Roche lobes is proportional to the orbital separation $a$. Usually, in computing binary evolution the departure from strict sphericity is ignored. The radius of a sphere $R_{RL}$ corresponding to star 1, with a volume equal to that of the lobe is usually approximated by the expression due to \citet{Eggleton}

\begin{equation}
R_{RL}= a\; \frac{0.49 q^{2/3}}{0.6  q^{2/3} + \ln{(1+q^{1/3})}}\; 0 < q < \infty
\end{equation}
\\ \noindent where $q=M_{1}/M_{2}$ is the mass ratio. The orbital period  $P_{orb}$ is related to $a$ by the Third Kepler's Law as $P_{orb}= 2\pi\big(a^3/(G [M_{1}+M_{2}])^{1/2}$; where $M_i,\ i=1,2$ are the masses of the components and $G$ is the Gravitational Constant.

If $P_{orb}$ is long enough, a binary system will always remain detached and each star evolves as if they were isolated. On the contrary, if $P_{orb}$ is short enough, the most massive star evolves faster\footnote{Notice, however, that in the case of compact companions, the component that evolves faster is the normal star, despite of its mass value.} and will fill its Roche lobe. Since at $L_{1}$, there is no net force capable of balancing the pressure gradient of the sub-photospheric layers, hydrostatic equilibrium is no longer possible and the system undergoes a {\it mass transfer episode}. Hereafter, we will refer as {\it donor star} to the one that is transferring mass and {\it accretor} to its companion. The components of the system may go through configurations unreachable for isolated stars. These systems are usually called Close Binary Systems (CBS). We denote the quantities related to the donor star with subindex 2 meanwhile for the companion star, we employ subindexes 1 (for normal stars), NS or BH, depending on the context.

This paper will focus on binaries that sometimes reach a semi-detached configuration during their evolution. Of course it is possible that, as a consequence of the mass exchange, the companion star evolves and swells enough to fill its Roche lobe, reaching a contact configuration. The evolution during and after contact is too complex to be treated with 1D simulations, and therefore, this evolutionary stage is beyond the scope of this work and will not be further discussed.

As stated above, understanding stellar evolution in binary systems is essential for accounting for a variety of astrophysical objects. A very important type of CBS is that formed by a normal star together with a NS. NSs are extremely compact objects with radii of tens of kilometers and masses up to $\approx 2M_{\odot}$. Their interior densities are $\approx 10^{15}\ g\ cm^{-3}$ \citep{Lattimer}. One of the most remarkable characteristics of magnetized NSs is that they support pulsar phenomenology, at least in zeroth order.


There are two well-known families of eclipsing binary millisecond pulsars in which one of the components is a normal star. Both have $P_{orb} \lesssim 1$~d. Black widows (BWs) have very light donor stars, with masses of $M_{2} \ll 0.1\; M_{\odot}$ whereas Redbacks (RBs) are around ten times more massive, with $0.1\; M_{\odot}\; \leq M_{2} \lesssim 0.4\; M_{\odot}$. Because of the association with the names given by the discoverers, they are now usually collectively termed as {\it spider pulsars} \citep{Roberts_2013}. It is widely believed that the donor stars first transfer mass to the NS, recycling their spin, and later the fastly spinning NS irradiates and evaporates the donor (see below for further details). This resembles the behaviour of the above-cited (black widow) spiders that, after mating, the female eats the male, but on a cosmic scale. Australians identified the higher $M_2$ group, and gave them the ``red back'' name corresponding to the Australian spider relative to the former black widow, an American one.


If a binary system including a NS is detached, it may be detected as a binary pulsar (in the sense of having a normal star companion). On the contrary, if it is semi-detached, it will be an X-ray source. If the companion star is a low (high) mass star, it is called Low (High) Mass X-ray Binary, LMXB and HMXB, respectively. Here, we are interested in LMXBs, and the class of spiders related to them.

One of the fundamental problems of binary evolution is related to the reaction of the donor star to the onset of mass transfer. Mass transfer may be a stable or unstable process depending on the ability of the donor to adjust itself to the change of the Roche lobe size due to mass exchange. Usually, stability is favoured if donors have a radiative envelope since they tend to contract. On the contrary, if the envelope is convective, the star swells as it transfers mass, which tends to destabilize the former process. It is also important that in the cases of interest, the donor star is not much more massive than its companion. If this were not the case, the size of the lobe would change too fast for the donor to be able to adjust to the lobe’s size. Moreover, if the timescale of mass accretion is shorter than its thermal timescale, the companion star cannot release the gravitational energy fast enough; thus, the system gets in contact. If the mass transfer rate evolves stably, it is possible to compute the evolution of the system. However, if it is unstable, the mass transfer will grow, and the system will enter the {\it common envelope} stage \citep{common_envelope}. Common envelope evolution is a very difficult issue, beyond the scope of the present paper. Some progress has been recently achieved in this problem \citep{Common}.


A mass transfer process may be conservative, in which the mass and angular momentum of the binary remain constant\footnote{Notice that, strictly speaking, in the case of short period CBSs, conservative evolution is impossible since Gravitational Radiation is non-negligible in this context.}. But non-conservative situations may occur. This mainly determines the evolution of the orbit of the system. 

It is usually considered that tidal effects \citep{hut} are so efficient that synchronization and circularization are instantaneous, and therefore the orbit remains circular over the entire evolution of the system. Relaxing these simple assumptions makes the numerical models largely increase their complexity.


Usually, and in the simplest form, the mass transfer process is described by two free parameters. $\beta$ is the fraction of mass transferred by the donor, and retained by the companion ($0\leq\beta\leq1$), and $\alpha$ is the specific angular momentum of the matter lost from the system in units of the specific angular momentum of the companion. Often, it is assumed that $\beta= 0.5$, and $\alpha=1$; see \citet{PRPh02} for further details. A NS can accrete up to a rate known as the {\it Eddington rate} $M_{Edd}= 2\times 10^{-8}\; M_{\odot}/yr$. The mass transfer rate $\dot{M}$ is usually expressed as \citep{Ritter_Mdot}

\begin{equation}
\dot{M}_{2}= -\; \dot{M}_{0}\; \exp{\big[\big(R_{2}-R_{RL}\big)/H_p}\big] \label{Eq:Ritter_mdot}
\end{equation}
\\ \noindent where $\dot{M}_{0}$ is the mass transfer rate for a donor star with radius $R_{2}$ that just fills its Roche lobe, and $H_{p}$ is the pressure scale height at the donor's photosphere.


Another ingredient of key relevance is {\it magnetic braking}. Suppose the donor star has a magnetic field attached to a convective envelope and suffers mass loss. In that case, this material corotates with the star along the field lines up to distances of a few stellar radii, slowing down its rotation. Since rotation is coupled to the orbit by tidal effects, the system's orbit is affected by this process. Remarkably, the strength and the functional dependence of braking remain considerably uncertain. The most used, ``standard law'' has been presented in \citet{1981A&A...100L...7V}. However, other prescriptions have been suggested by \citet{2019MNRAS.483.5595V} and \citet{2019ApJ...886L..31V}. The results presented and discussed below are based on the standard prescription. 

\section{binary systems with a neutron star companion: standard evolution} \label{Sec:conNSs}

Usually, in the studies of LMXBs it is assumed that the evolution of the systems is entirely due to the mass transfer from the donor toward the NS component, and the consequent reaction of the orbit to it. It is well-known that an accretion disk \citep{Pringle} must be formed around the NS. However, for simplicity, the effects due to the presence of the disk are neglected. Moreover, the material that falls onto the NS produces X-rays and these must lead to the irradiation of the donor star. This is also ignored in general. Because of this reason, this family of standard models may also be referred to as {\it non-irradiated models}.

\begin{figure*}[t]
\centerline{\includegraphics[width=0.90\textwidth,angle=0]{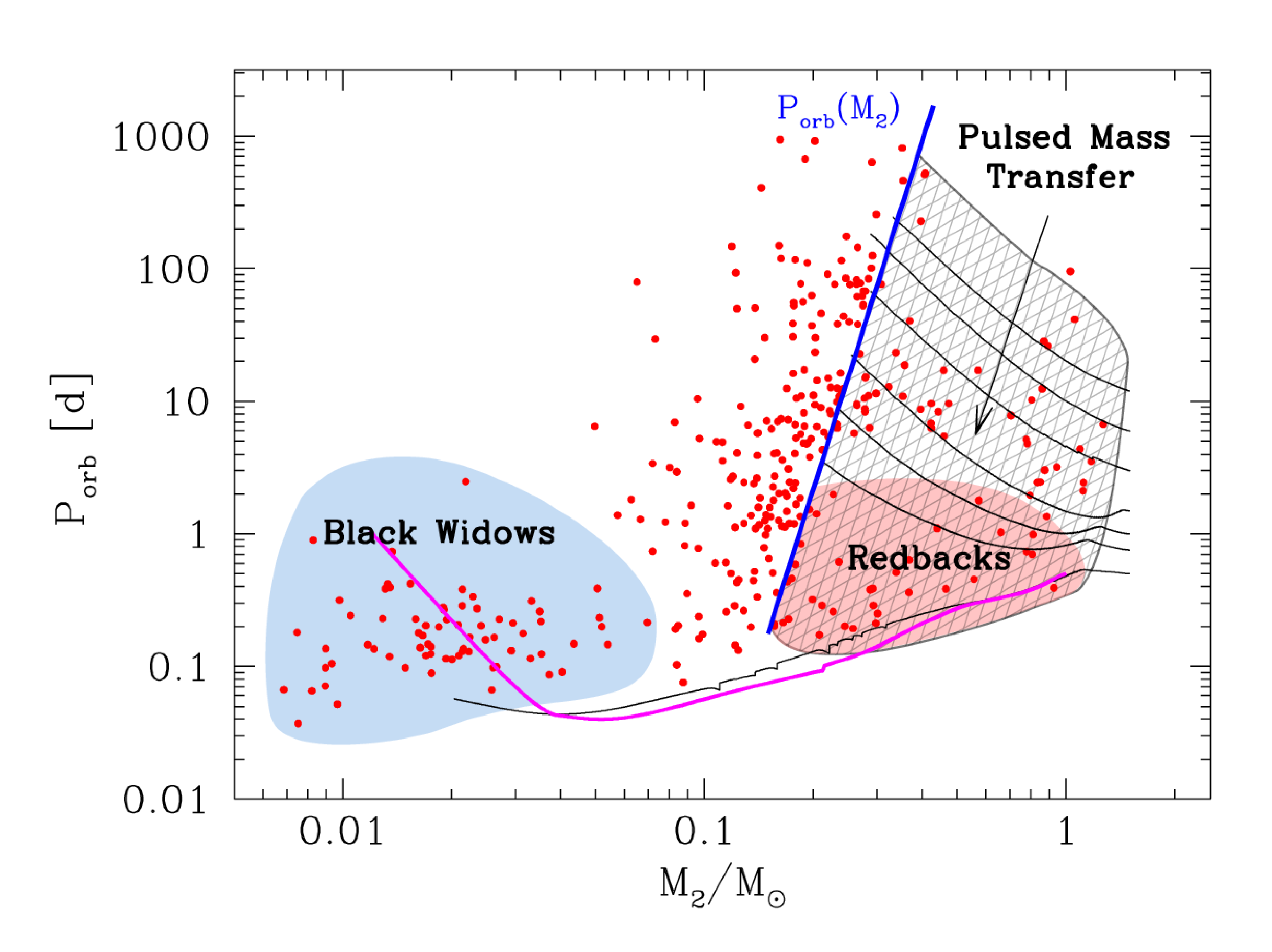}}
 \caption{The donor mass vs. orbital period plane for CBSs containing a pulsar. Minimum mass vs. orbital period for pulsars belonging to binary systems with quasi-circular orbits (with eccentricity $e\leq 10^{-3}$) were selected from the ATNF database \citep{ATNF}. The steep relationship of $P_{orb}(M_{2})$ shown with a thick blue line is that of \citet{Masa_periodo}. Also, with black thin lines we included a set of evolutionary tracks corresponding to solar composition objects with masses for the donor star and the NS of $M_{2}= 1.50\;M_{\odot}$, and $M_{NS}= 1.40\;M_{\odot}$ respectively, for initial orbital periods of $P_{orb}= 0.50, 0.75, 1.0, 1.5, 3.0, 6.0,$ and $12.0$~d ordered from bottom to top. The thick magenta line corresponds to a model with $M_{2}= 1.50\;M_{\odot}$, $M_{NS}= 1.40\;M_{\odot}$, and $P_{orb}= 0.50$~d with moderate evaporation (see text (\S~\ref{evaporando}) for further details). The regions corresponding to BWs (evaporation of the companion) and RBs (accretion and recycling) are highlighted. Also, the region corresponding to pulsed mass transfer due to IFB is indicated. Notice that the latter and RBs region overlap.}
\label{Fig:ATNF}
\end{figure*}

Non-irradiated LMXBs models predict the occurrence of long-standing mass transfer episodes on timescales of a few Gyr. During this stage, the objects are expected to be visible as LMXBs, while after detachment, they can be detected as binary pulsars. For the case of orbital periods $P_{orb} \gtrsim 0.1$~d, there is a well-known relation between $P_{orb}$ and the donor mass $M_{2}$  (\citealt{Tau_Savo}; \citealt{Masa_periodo}) hereafter defined as $P_{orb}(M_{2})$. In Figure~\ref{Fig:ATNF}, we show the donor mass vs. orbital period plane for binary pulsars with well-circularised orbits (with an eccentricity lower than 0.001), taken from the ATNF database\footnote{\url{www.atnf.csiro.au/research/pulsar/psrcat/}}  \citep{ATNF}. These data correspond to the {\it minimum mass} of the objects. We include the $P_{orb}(M_{2})$ relation given in \citet{Masa_periodo}. Also, we included a set of evolutionary tracks corresponding to solar composition donor star and NS with initial masses of $M_{2}= 1.50\;M_{\odot}$, and $M_{NS}= 1.40\;M_{\odot}$ respectively, for initial orbital periods of $P_{orb}= 0.50, 0.75, 1.0, 1.5, 3.0, 6.0,$ and $12.0$~d. We have also included a track for the case of $M_{2}= 1.50\;M_{\odot}$, $M_{NS}= 1.40\;M_{\odot}$ and $P_{orb}= 0.50$~d with moderate evaporation (see below \S~\ref{evaporando}).

Notice that we know the minimum mass values for the observed binary pulsars. Many of them fall inside the highlighted region of Figure~\ref{Fig:ATNF} in which a continuous mass transfer (corresponding to an LMXB) is expected. This is in sharp contradiction with observations. Standard models predict the occurrence of binary pulsars in the surroundings of the $P_{orb}(M_{2})$ relation. Evidently, the standard theory fails to account for all the available observations.


As stated above, all these results are based on the standard prescription for magnetic braking and without irradiation. For an analysis of the effects of other treatments of magnetic braking, we refer the reader to \citet{Maite_Frenados}.

\section{binary systems with a neutron star companion: Non-standard evolution} \label{sec:Ifeedback}

\subsection{The Irradiation feedback}

As stated above, the standard model of LMXBs evolution neglects the effects induced by the X-ray illumination and evaporation of the donor star; see, e.g., \citet{PRPh02}. This illumination yields a minor effect if the donor has a radiative envelope. However, if it has a thick enough outer convective envelope with a column density $\Sigma > 10^{6}\ g\ cm^{-2}$, the consequences of considering the incident radiation are not negligible. Indeed, if incident X-ray radiation is strong enough, the illuminated hemisphere is partially inhibited from releasing energy emerging from the stellar interior. This can be viewed as if the stellar surface were effectively smaller \citep{Buning-Ritter}.

In order to emit X-rays, the NS must undergo mass accretion. Compared with the mass transfer rate it would have without considering irradiation, irradiation feedback (IFB) makes the star undergo an enhanced $\dot{M}$ for a while. Sometime later, the donor is unable to sustain this $\dot{M}$ value; and thus recedes and detaches. Subsequently, nuclear shell burning forces the star to swell again, entering the semi-detached configuration and re-establishing mass transfer. Thus, IFB makes the mass transfer process occur not continuously, but in a sequence of long pulses \citep{Pulsos_Mdot}. When the system is transferring mass, it is observable as a LMXB, while when it is detached, it may be detectable as a binary pulsar. Thus, models without IFB predict a continuous, long-standing mass transfer episode, while irradiated models may undergo a pulsed mass transfer. This qualitative difference is crucial for accounting for some of the observed binary pulsars included in Figure~\ref{Fig:ATNF}. 

The accretion luminosity $L_{acc}$ radiated by the NS is

\begin{equation}
L_{acc}= \frac{G M_{NS} \dot{M}_{NS}}{R_{NS}},  
\end{equation}

\noindent where $M_{NS}$, $R_{NS}$, and $\dot{M}_{NS}$ are the mass, radius and accretion rate of the NS, respectively. The photon flux incident onto the donor star $F_{ifb}$ that participates in the process is given by $F_{ifb}= \alpha_{ifb}\ L_{acc}/(4\pi\ a^{2})$, where $\alpha_{ifb}$ is a parameter ($0 < \alpha_{ifb} < 1$). \citet{Hameury-Ritter97} have presented a simple method to consider IFB that we have included in our stellar code \citep{BDV_Codigo}. With this tool, we have computed a series of models to explain the existence of binary pulsars. We have presented these results in a series of papers  \citet{BDVH_BlackWidows,BDVH_Understanding,BDVH_QuasiRLOF,BDVH_RedbacksGC} and highlight the results below. 

It has been recently claimed that there is strong evidence for yet another family of spiders: the so-called ``Huntsman'' pulsars \citep{Huntsman} (another true spider). In that paper, the authors associate the occurrence of Huntsman spiders with a well-known evolutionary stage of low-mass stars: after exhausting its hydrogen core, a low-mass star develops a hydrogen shell burning and a very deep convective envelope that produces a two step profile. When shell burning reaches the edge of the outer step, it enhances its energy release, making the star react by contracting and detaching for a while (usually a few tens of millions of years). Then, due to the outwards shell burning, the star inflates again. This behaviour is the usual explanation for the occurrence of the ``red clump'' observed in many open and globular clusters; see, for example, \citet{RedClump}.

In the context of binary evolution, the contraction due to nuclear burning quoted above is responsible for the detachment of the binary and the possibility of observing the system as a binary pulsar. \citet{Huntsman} claimed this is explained without any unusual assumption. This is a plausible hypothesis. However, there is no way to account for the existence of the many binary pulsars shown in Figure~\ref{Fig:ATNF} for which standard models predict a long-standing mass transfer episode. Although not really proven to be the responsible for the observed behaviour, the IFB provides a very reasonable explanation. We should remark that both phenomena (IFB and evaporation) are already present in the suite of models presented in our papers cited above, especially in \citet{BDVH_Understanding}, although this very particular episode of nuclear burning was not discussed there. There is no contradiction between the mechanism proposed by \citet{Huntsman} for the existence of Huntsman pulsars and the occurrence of IFB. The latter will make a small difference for the evolution of the Huntsman group, since IFB does not affect the region at the base of the convective envelope deep below the surface.

In any case, it should be stressed that the whole modeling of this kind of system is far from complete. One of the main uncertainties is related to the fraction $\alpha_{ifb}$ of X-ray irradiation that acts in the IFB process. If it is small, there will be many mass transfer peaks; if it is larger, fewer peaks will occur (See Fig.~2 of \citealt{BDVH_Understanding}).

\begin{figure}[t]
\centerline{\includegraphics[width=0.40\textwidth,angle=270]{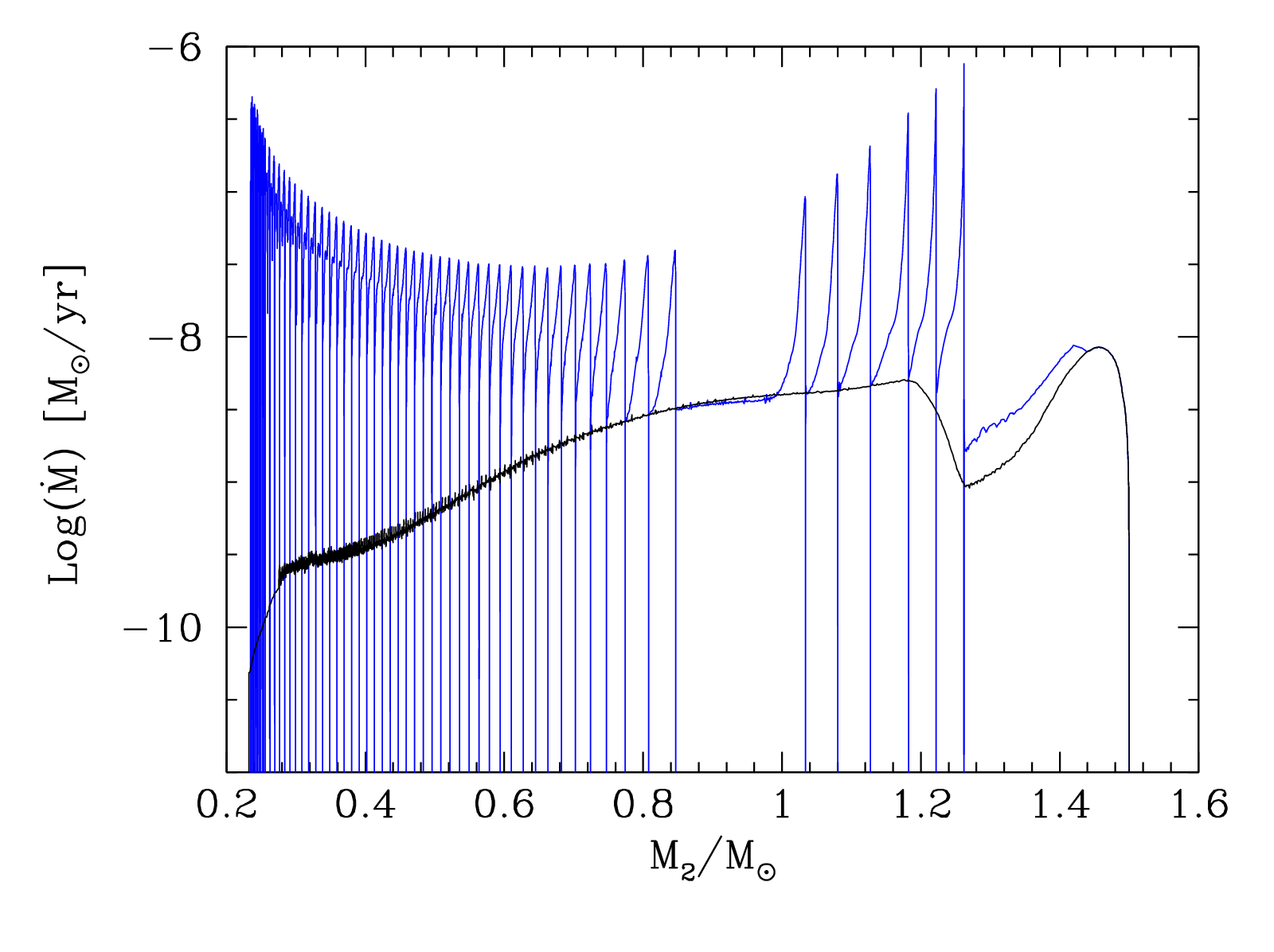}}
 \caption{Mass transfer rate as a function of donor mass for a solar composition donor star and NS of $M_{2}= 1.50\;M_{\odot}$, and $M_{NS}= 1.40\;M_{\odot}$ respectively, for an initial orbital period of $P_{orb}= 1.0$~d. Blue lines correspond to an irradiated model with $\alpha_{ifb}= 0.10$, whereas black lines correspond to a non-irradiated model. \label{Fig:Mdot_vs_m2}}
\end{figure}

\begin{figure}[t]
\centerline{\includegraphics[width=0.40\textwidth,angle=270]{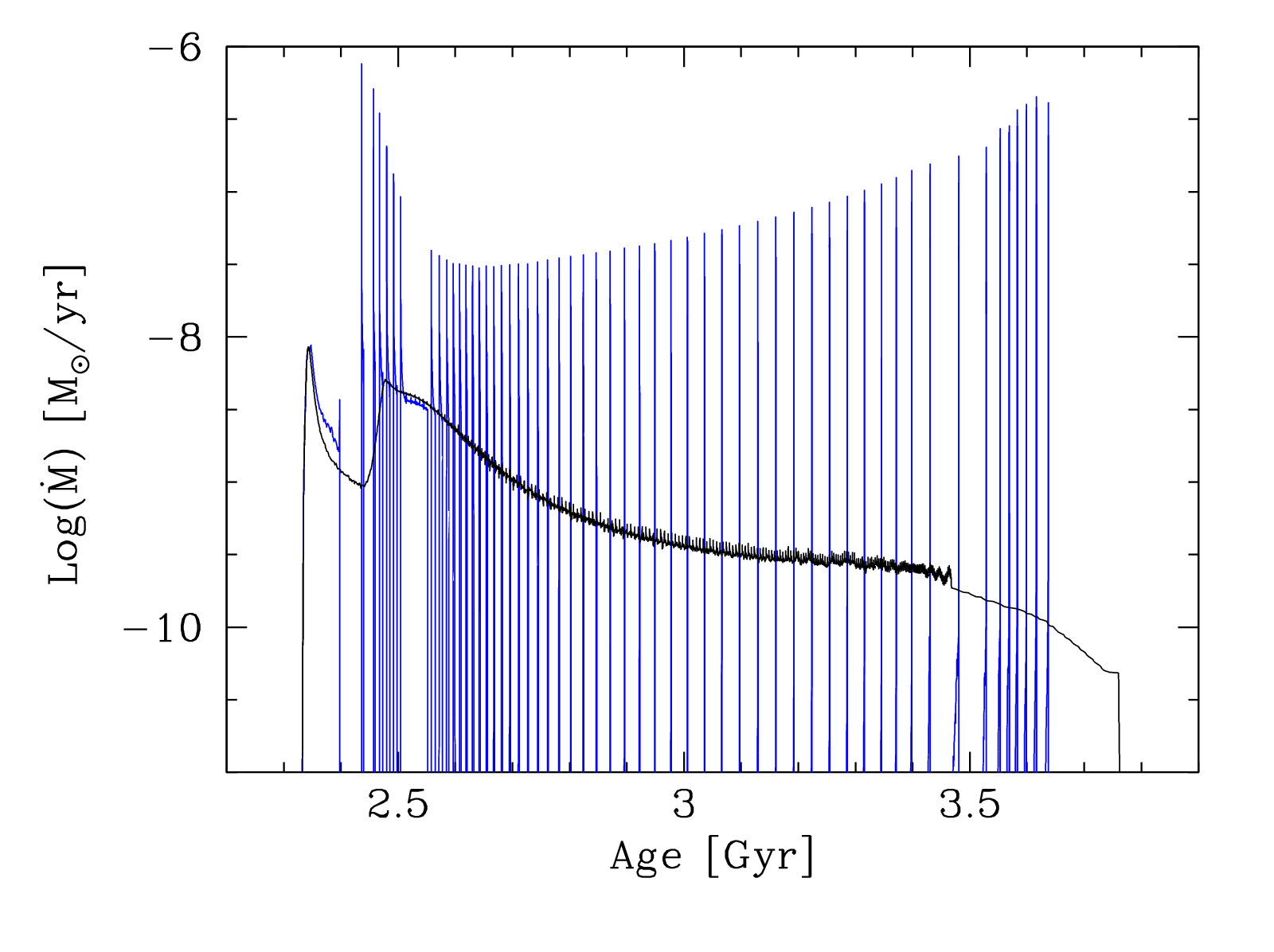}}
 \caption{Same models as Figure~\ref{Fig:Mdot_vs_m2}, but for mass transfer rates as function of time. Notice that, in the case of models with IFB,  most of the time the system remains detached. \label{Fig:Mdot_vs_t}}
\end{figure}

In Figure~\ref{Fig:Mdot_vs_m2}, we show the evolution of the mass transfer rate as a function of the donor mass for the particular case of a system with solar abundances donor star and NS whose initial conditions are masses of $M_{2}= 1.50\;M_{\odot}$, $M_{NS}= 1.40\;M_{\odot}$ and $P_{orb}= 1.0$~d. Blue lines in the figure correspond to an irradiated model with $\alpha_{ifb}= 0.10$, whereas black lines correspond to a non-irradiated model. For the same models, in Figure~\ref{Fig:Mdot_vs_t}, we show the evolution of $\dot{M}_{2}$ as a function of time. From these figures, we note that considering IFB or not is of key importance in interpreting the occurrence of binary pulsars/accreting systems. IFB makes the systems to be detached most of the time, allowing the observation of the pulsars. Quite on the contrary, non-irradiated models do not allow that possibility and therefore lead to a conflict with the objects located to the right of the $P_{orb}(M_{2})$ relation presented in Figure~\ref{Fig:ATNF}. 

Perhaps one of the most intriguing characteristics of the models with IFB is that the long-term behaviour is not strongly dependent on it. For example, the evolution of the systems in the $P_{orb}$~vs.~$M_{2}$ plane (See Figure~\ref{Fig:ATNF}) relation is almost unaffected \citep{Ritter_review}. Because of this fact, and considering that computing irradiated models may be very time-consuming, IFB has been usually ignored in the calculations, although it is important for the interpretation of the data and the understanding of the systems, as stated above. 

\begin{figure}[t]
\centerline{\includegraphics[width=0.40\textwidth,angle=0]{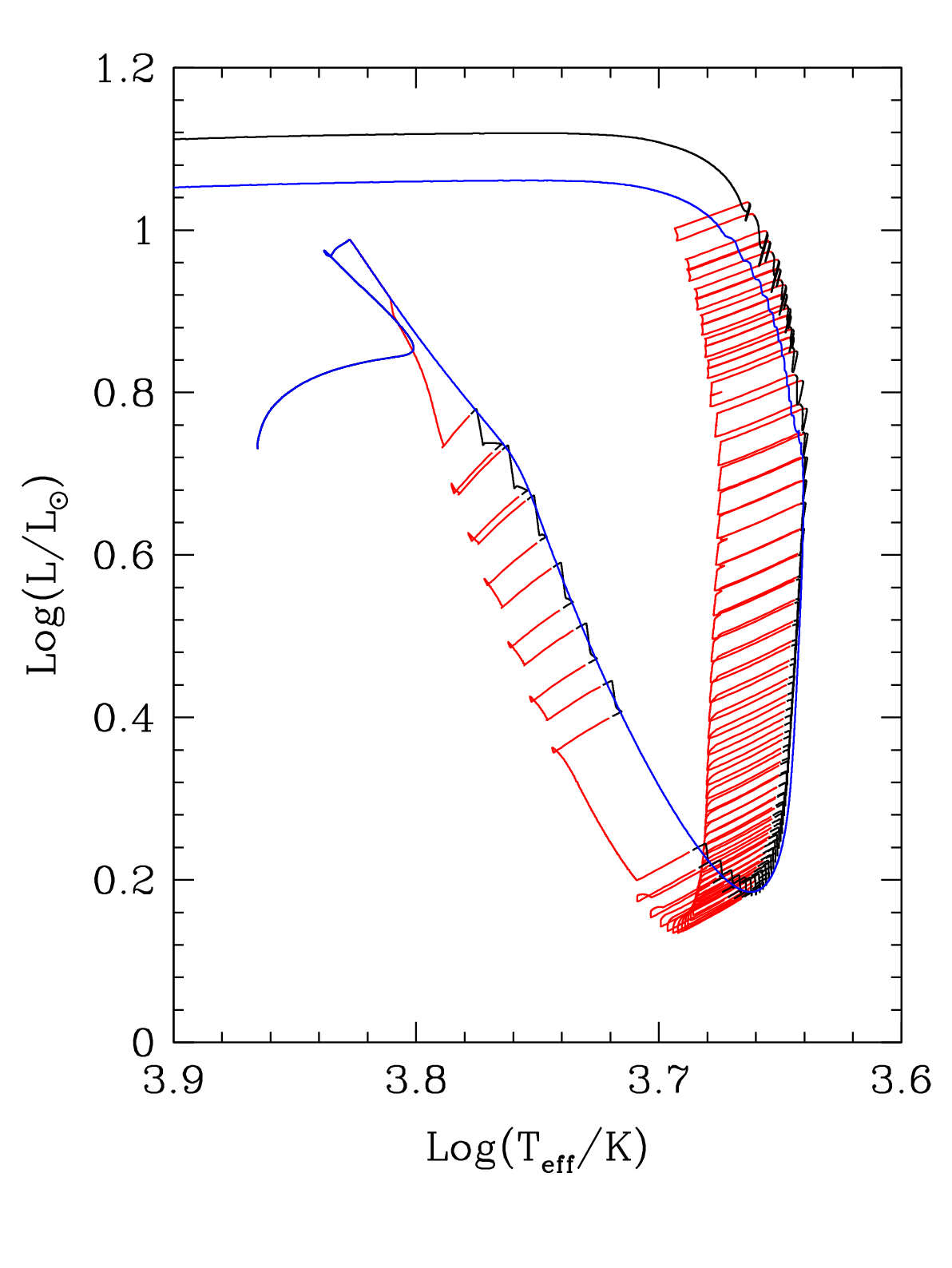}}
 \caption{Hertzsprung-Russell Diagram for the same models included in Figures~\ref{Fig:Mdot_vs_m2}-\ref{Fig:Mdot_vs_t}. The track corresponding to non-irradiated models is presented in black line. The irradiated model is depicted in blue for detached stages, whereas semi-detached stages with mass transfer are denoted with red lines. \label{Fig:HRI}}
\end{figure}

\begin{figure}[t]
\centerline{\includegraphics[width=0.40\textwidth,angle=0]{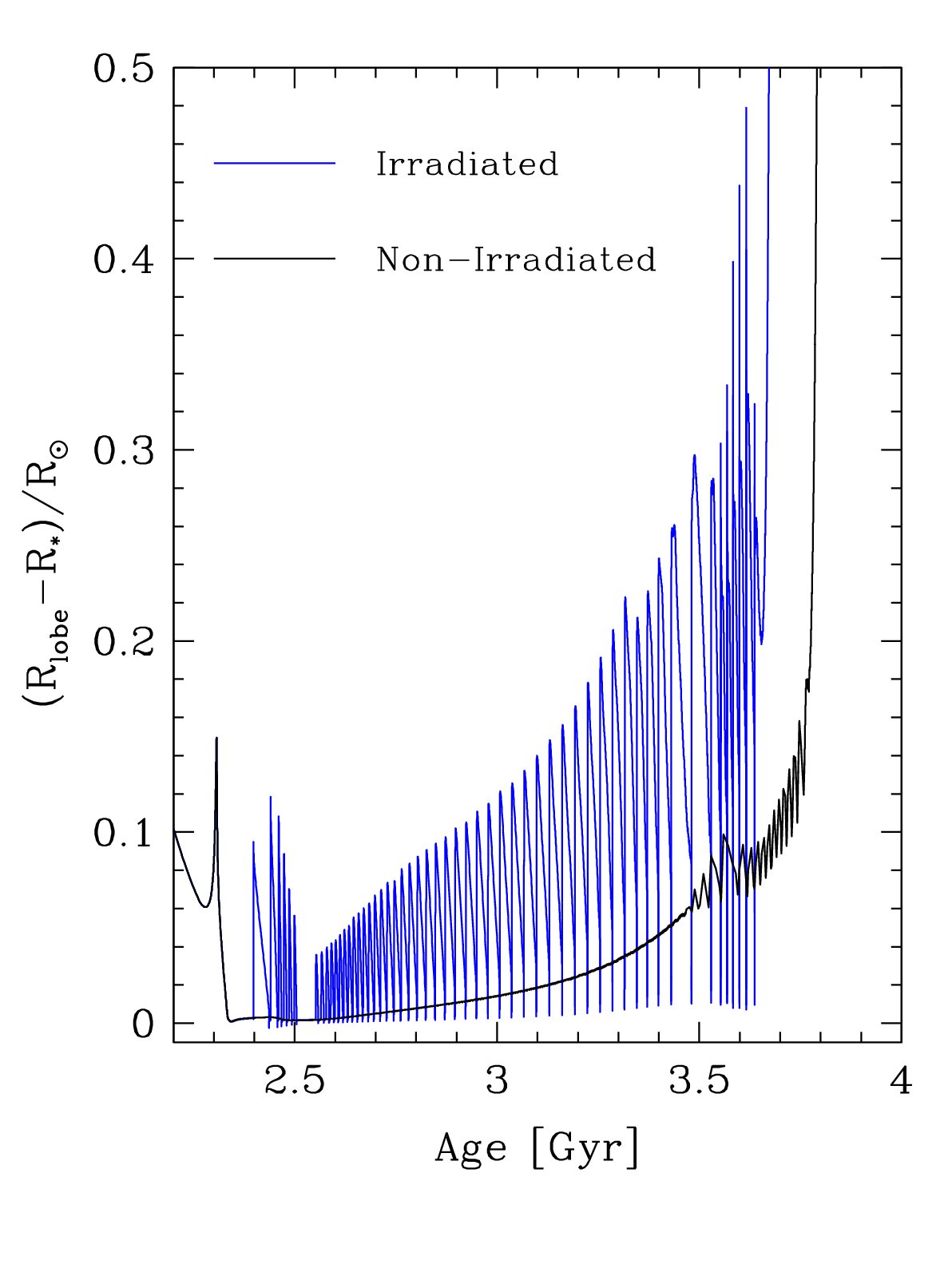}}
 \caption{The difference between the radius of the Roche lobe and that of the star for the same models included in Figures~\ref{Fig:Mdot_vs_m2}-\ref{Fig:Mdot_vs_t}. \label{Fig:RadiosI}}
\end{figure}

To show this feature in more detail, we show in Figure~\ref{Fig:HRI} the track in the Hertzsprung-Russell diagram (HRD) for the models included in Figures~\ref{Fig:Mdot_vs_m2}-\ref{Fig:Mdot_vs_t}. There, the
portion of the irradiated model in which there is mass transfer is denoted with red lines. Non-irradiated models (in black) and irradiated detached models have very similar tracks. Furthermore, in Figure~\ref{Fig:RadiosI} we show the difference between the lobe size and stellar radii, which determines $\dot{M}_{2}$ (see Equation~\ref{Eq:Ritter_mdot})

We believe that these models as promising for interpreting the occurrence of the whole binary pulsar population, particularly the spider systems. However, we should stress that they include severe approximations. The models assume a {\it point source} X-ray irradiation coming from the NS surface, and the presence of an accretion disk is ignored. Thus, the shortest timescale present in the models is the thermal Kelvin-Helmholtz scale of the donor star envelope. Any faster variation of the aspect of the systems cannot be interpreted in the frame of these models. Moreover, part of the irradiation may be coming from the accretion disk surrounding the NS, and it may partially eclipse that coming from the NS surface. This might even partially preclude the occurrence of pulsed mass transfer.

In the context of IFB models, in \citet{Leonela} we have relaxed the hypothesis of instantaneous synchronization and circularization in between mass transfer pulses to investigate the origin of the large value of $\dot{P}_{orb}= -(3.50 \pm 0.12) \times 10^{-9}\ s\ s^{-1}$ detected by \citet{Crawford} in the RB system PSR~J1723-2837. Such a period derivative is far larger than expected for an instantaneously synchronized detached system. It has been concluded that pulsed mass transfer provides a promising context to provide a plausible explanation for the origin of large-period derivatives. The observation of additional cases would be important for the whole picture.

\subsection{The evaporation phase} \label{evaporando}

Another important phenomenon considered in this kind of systems is evaporation. Due to the pulsar irradiation impinging the donor star, a strong wind would act onto the donor, regardless of whether the system is well detached. This has been proposed to account for the observed characteristics of the first detected BW. A high-energy version of an ordinary comet is the emerging picture in black widow systems \citep{BW_Original}.

Evaporation makes the system lose mass and angular momentum. This leads to the increase of $P_{orb}$, widening the system's orbit. In the plane $M_{2}$~vs.~$P_{orb}$ presented in Figure~\ref{Fig:ATNF}, the evolutionary track of the system with evaporation bends upwards, reaching the observed relatively long orbital periods. Notice that models without evaporation are unable to cover the observed range of periods corresponding to BWs.

The evaporation rate is usually described in a parametric way, as proposed by \citet{Rate_evap} 

\begin{equation}
\dot{M}_{evap}= -\; \frac{\alpha_{evap}}{2 v^{2}_{2,esc}}\; L_{PSR} \bigg(\frac{R_{2}}{a}\bigg)^{2}
\end{equation}
\\ \noindent where $L_{PSR}$ is the pulsar spin-down luminosity, $v_{2,esc}$ is the escape velocity at the donor star photosphere, and $\alpha_{evap}$ is an efficiency parameter. The stronger the evaporation, the higher the reached $P_{orb}$. In Figure~\ref{Fig:ATNF}, we have included the evolutionary track (denoted with a magenta line) for the case of $M_{2}= 1.0\;M_{\odot}$, $M_{NS}= 1.40\;M_{\odot}$, $P_{orb}= 0.50$~d  and $\alpha_{evap}\ L_{PSR}= 0.15\ L_{\odot}$.


\subsection{Do BWs evolve from RBs?}

A careful inspection of Figure~\ref{Fig:ATNF} indicates that the evolutionary tracks that reach the BW region previously go through the RB zone. However, not all RBs are expected to evolve to BWs, since for the case of large enough $P_{orb}$ values, they end as detached systems not reaching the evaporation stage. This result, presented in \citet{BDVH_Understanding}, is in sharp contrast with the model proposed by \citet{2013ApJ...775...27C} who considered very high evaporation rates. Indeed, one of the consequences of the scenario presented by \citet{2013ApJ...775...27C} is that the donor star should be detached from its Roche lobe; i.e., the filling factor should be very low. On the contrary, the model that proposes IFB as the physical phenomenon that leads to the occurrence of RBs predicts the opposite situation. In this case, the Roche lobe is only slightly underfilled, as shown in \citet{BDVH_QuasiRLOF} and \citet{DVBH_Formatio_Redbacks}. In fact, in Table 2 of that paper we listed the filling factors obtained from observations, and their tendency to high values seem to favor our models over that of Chen et al. (2013). 

Even though much work remains to be done to reach a more compelling description of these fascinating objects, the IFB and later evaporation phenomena \citep{Yep, Trucho} proved potentially able to address all the systems without any obvious contradictory prediction. This is why we insist in their testing and further elaboration.

\subsection{Maximum masses and collapses to black holes}

Given that the typical evolution times are ~several Gyr, even at Eddington-limited accretion rates with intermittent behaviour, it is natural to wonder whether the NS formed and accreting material could achieve high masses, or even be forced to collapse to BH when the maximum mass of the sequence is reached. Within the treatment of the accretion problem, it is quite difficult to address these questions, mainly because the tracks assume a constant value of the irradiation $\alpha_{irr}$ and other effects (like the so-called {\it radio ejection}; see, e.g.,  \citealt{Burderi}) were ignored. A self-consistent complete calculation is not yet available, but the problem is important enough to motivate a brief appraisal here.

\begin{figure}[t]
\centerline{\includegraphics[width=0.50\textwidth,angle=0]{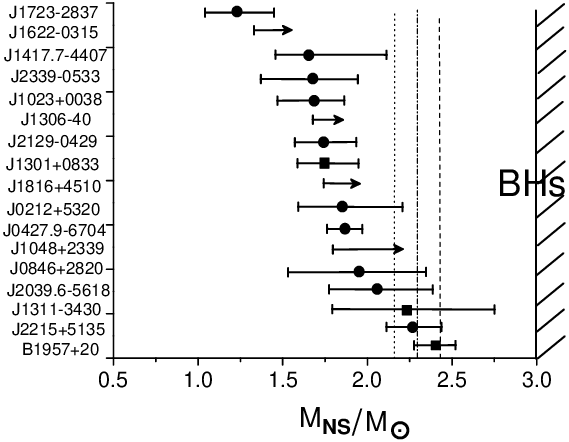}}
 \caption{Masses of spider NSs determined with optical lightcurves \citep{HorvathChina}. \label{Fig:MChina}}
\end{figure}

Masses of NS in these RB-BW systems show an intriguing duality associated to their extraction method, i.e., they come out different if optical light curve fitting or gamma-ray eclipses are employed \citep{Clark}. The eclipse analysis gives lighter NS masses, and the exact reasons (and, therefore, the masses) remain unclear. The maximum mass reported is $2.35 \pm 0.17 \, M_{\odot}$ (\citet{Romani}), for the spider pulsar PSR J0952–0607. It suggests by itself that the accretion history may push the mass value up close to the (theoretically unknown) Tolman-Oppenheimer-Volkoff (TOV) limit value. A collection of reported masses and a discussion of this idea has been presented by \citet{HorvathChina}; see also \citet{NSmasiva}. The situation can be appreciated in Fig.~\ref{Fig:MChina}.

A simple, but not proven, interpretation is that the NS systems can become very heavy depending on their detailed history. Up to $\sim 1 \, M_{\odot}$ may be accreted, according to \citet{Linares} calculations, suggesting that the lightest conceivable NS at the beginning of the accretion stage could go over the $2 \, M_{\odot}$ benchmark. This has been suggested by \citet{HorvathBenvenutoCSQCD} and certainly merits a further study. 

An extreme version of an efficient accretion process could indeed push some NS over the TOV limit, even if the latter happens to be as high as $\sim 2.5 M_{\odot}$ \citep{nosotros}. Moreover, and in the same line of possible suggestions, if the ``kick'' received by the new formed BH at the edge of the TOV value is not large, the system would remain bound and feature a low-mass black hole with $\sim 3 \, M_{\odot}$. A detection of CBSs of these type would be very interesting and highlight a new form of producing objects well inside the ``mass gap'' which may not really exist as such \citep{Lucas}.

\section{systems with a black hole companion} \label{Sec:conBHs}

Another interesting kind of CBS is that in which the compact companion to the donor star is not a NS but a BH. As in the case of systems with NS companions, the transferred mass forms an accretion disk around the compact companion that may accrete a fraction of it while the rest is lost from the system. In this context, it is expected that the occurrence of long-standing mass transfer episodes. Here, the only possible X-ray emitter is the accretion disk surrounding the BH. Such a disk is unable to sustain the emission for timescales comparable to the thermal timescale of the donor. This precludes the occurrence of the mass transfer cycles as those expected for the case of NS companions \citep{Ritter_review}.

In \citet{V404Cyg} we have investigated the case of V404~Cyg, a LMXB with a BH that has suffered outbursts in 1938, 1989, and 2015. Many data are available for this system, making it particularly interesting to study with detailed models: the masses of the components, the orbital period, the donor star spectral type, and chemical composition are well estimated. The donor star has heavy element abundances two times higher than those in our Sun \citep{compo_V404}. \citet{Polacos_bh} have studied V404~Cyg, finding that on average, the BH should be accreting mass at a rate of $\dot{M}_{BH}=4.0 \times 10^{-10}\ M_{\odot}\ y^{-1}$. 

To study the formation of V404 Cyg, as usual for this kind of CBS, we assumed that the BH was initially non-rotating \citep{BH_norota}. As the BH gains mass, its spin parameter, defined as $a^*\equiv cJ_{BH}/GM_{BH}^2$, increases following:

\begin{equation}\label{BHspinparameter}
    a^*=\left( \frac{2}{3} \right)^{\frac{1}{2}} \frac{M^0_{BH}}{M_{BH}} \left\{ 4-\left[ 18\left( \frac{M^0_{BH}}{M_{BH}} \right)^2 -2\right]^{\frac{1}{2}} \right\},
\end{equation}
\\ \noindent where $M^0_{BH}$ and $M_{BH}$ (with $M_{BH} < \sqrt{6}\ M^0_{BH}$ so, $0 < a^* < 1$) are the initial and present masses of the BH, respectively.

We found that most of the characteristics of V404~Cyg  are explained by the evolution of a system initially formed by a BH of 9~$M_{\odot}$ and a donor of 1.50~$M_{\odot}$. However, remarkably, this model has failed by far to reach the spin regime of the BH. For V404~Cyg, the most probable value is $a^* > 0.92$ \citep{V404Cyg_obs}. Indeed, the highest values of $a^*$ obtained from our (initially non-rotating) models were about one-third of the value deduced from observations.

The results presented in \citet{V404Cyg} for the case of V404~Cyg suggest that, at least in some cases, if initially the BH does not rotate, mass transfer in CBSs is not enough to speed up rotation to the observed regime.   

\section{Blue straggler stars as due to binary evolution} \label{sec:BSSs}

Another type of object, whose existence is currently believed to be (at least partially) due to binary evolution, is blue straggler stars (BSSs).

One of the most important predictions of stellar evolution is that the more massive the star, the faster the evolution. This fact has a well-known consequence: the HRD of star clusters shows the existence of a {\it turning point} that corresponds to the brightest stars still belonging to the Main Sequence.

\begin{figure}[t]
\centerline{\includegraphics[width=0.45\textwidth]{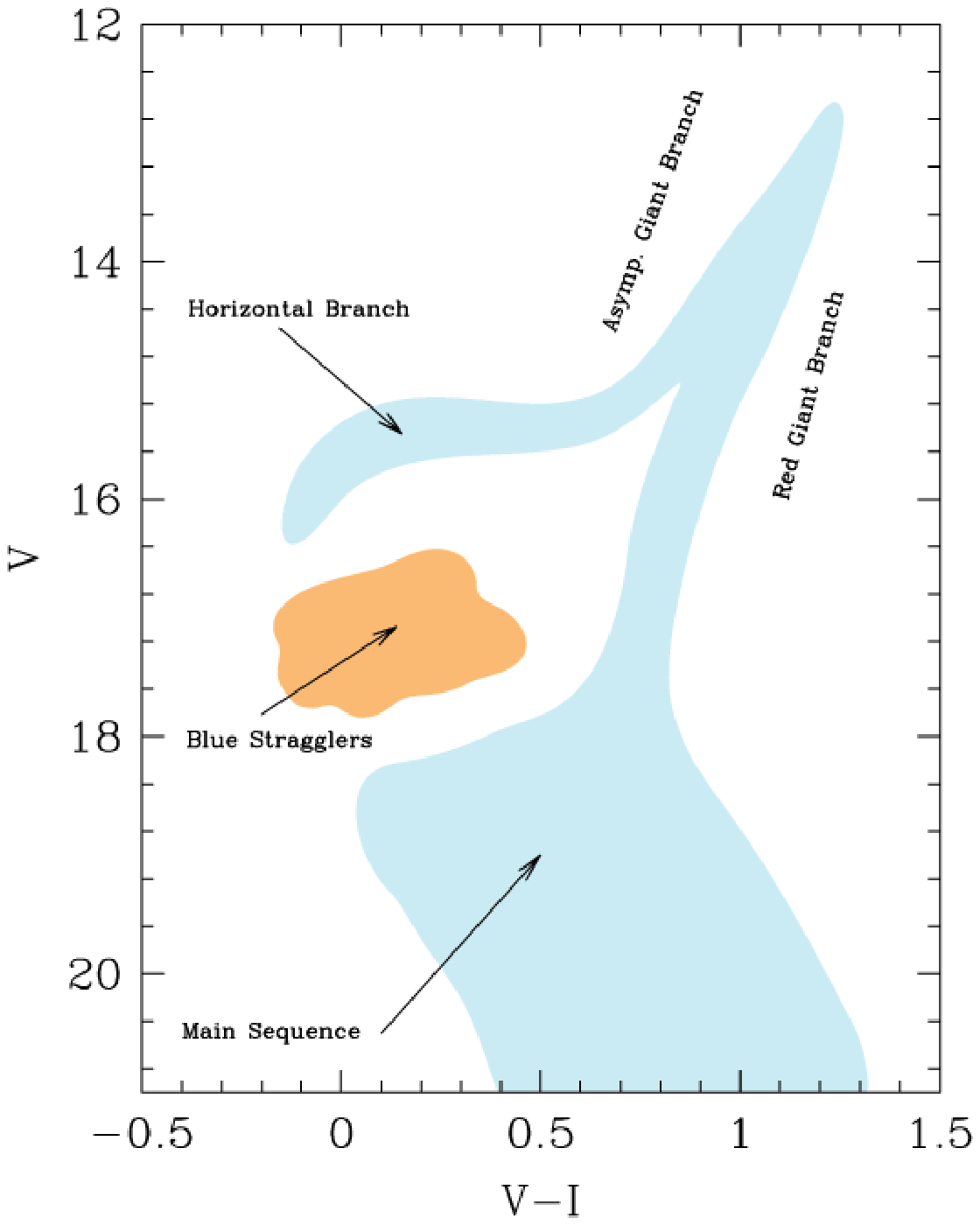}}
 \caption{Schematic color-magnitude diagram corresponding to the Messier~3 globular cluster. The population of stars is identified with a variety of evolutionary stages corresponding to single stars, described in \citet{1990sse..book.....K}. Blue Straggler Stars fall in between the turn off point at the top of the main sequence and the horizontal branch. They {\it cannot} be interpreted as due to single star evolution.  \label{Fig:Messier3}}
\end{figure}

In Figure~\ref{Fig:Messier3}, we show a schematic color-magnitude diagram of the globular cluster Messier~3 (in the Johnson filter system), showing the regions with a high density of objects (for a description of the characteristics of this particular globular cluster, see e.g., \citealt{Messier_3}). 

Single star evolution predicts that the low-mass stars that populate the globular cluster when they exhaust their hydrogen core, leave the main sequence band and evolve upwards along the red giant branch. Then, when the hottest (off-center) degenerate layers reach a temperature of $T \approx 10^{8}~K$, they suffer the helium flash, after which they settle for a long time on the horizontal branch (that is why it is detectable), undergoing stable core helium burning. Then, when the helium core is exhausted, the star evolves to the right, upwards along the asymptotic giant branch (up to this point, all these stages are denoted in the Figure~\ref{Fig:Messier3}) and finally suffers several thermal pulses and evolves bluewards very fast to become a white dwarf \citep{1990sse..book.....K}. The turn-off point corresponds to the most luminous stars still belonging to the Main Sequence. BSSs are in between the turn-off point of the Main Sequence and the Horizontal Branch.  

BSSs were first detected by \citet{Sandage} in the core of Messier~3. If all the stars of the cluster had been born simultaneously, to be brighter, they would have been more massive, and so, they should have had a faster evolution. Thus, BSSs are located in a region of the HRD impossible to be occupied by objects born as single stars. The solution to this well-known paradox is widely believed to be connected to the interaction of a pair of (or even three) stars.

One possible scenario is the mass transfer in a CBS. When the donor star fills its Roche lobe, it begins to transfer material to the companion accreting star. This, in turn, modifies the structure of the latter, becoming progressively more massive and brighter. In this way, the star resembles a younger object, reaching a region of the HRD near the Main Sequence but also above the turning point of the cluster it belongs, becoming a BSS. This mechanism was proposed long ago by \citet{McCrea_BSSs}. 

Another possibility proposed to explain the formation of BSSs is stellar collisions \citep{Colisiones}. In this case, two stars suffer a head-on collision in a way that the resulting object is more massive than they were any of them previous to the interaction. As a consequence, the structure of the resulting star accommodates to resemble a younger and more massive star settling in the HRD just at the BSSs region. Collisions are expected to be more frequent in a densely populated environment as provided by a globular cluster. However, remarkably, it has been claimed that collisions may be necessary in accounting for the BSS population of open clusters \citep{BSS_open_crash}.

\begin{figure}[t]
\centerline{\includegraphics[width=0.40\textwidth,angle=270]{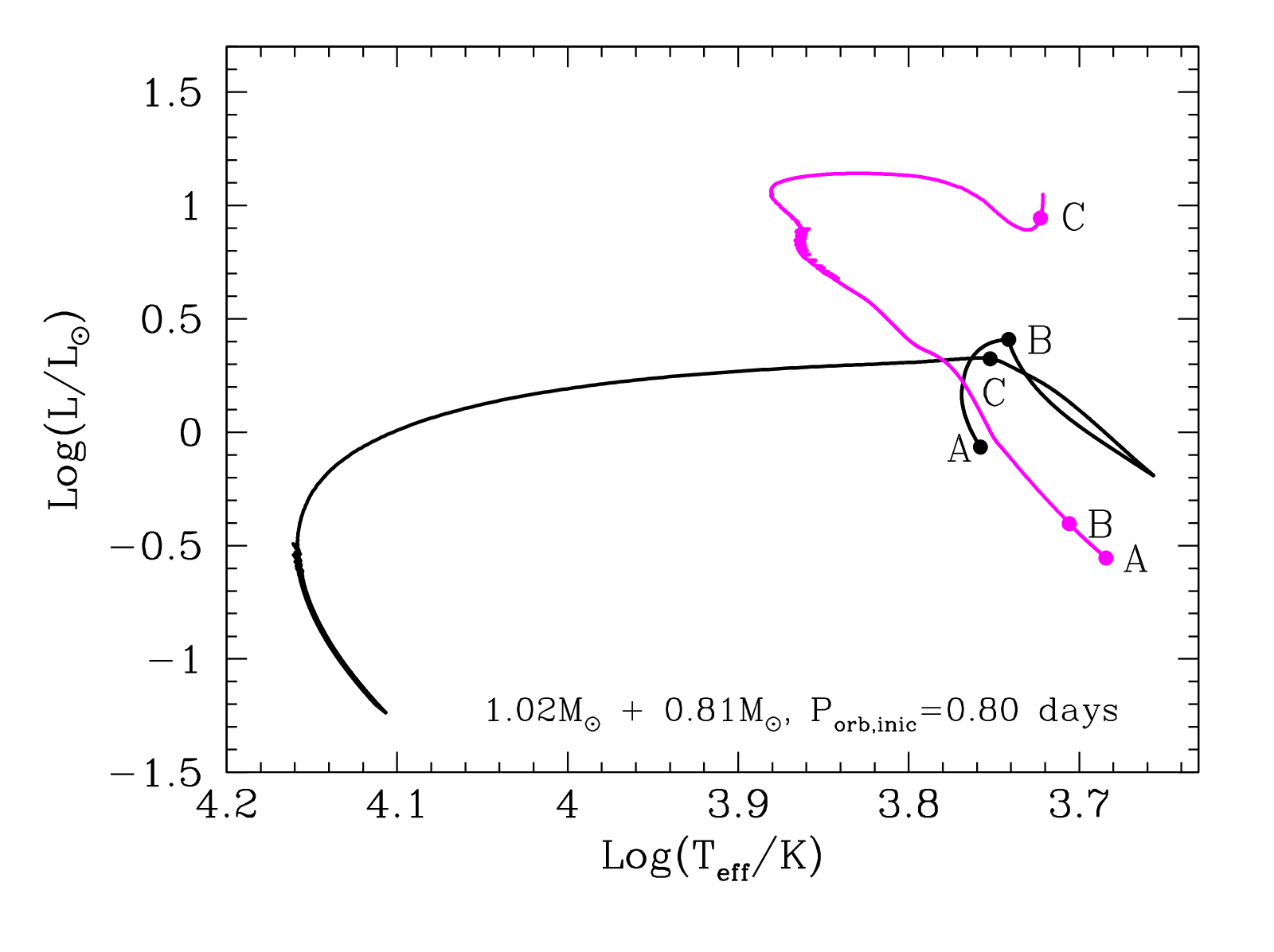}}
 \caption{The evolutionary tracks for the components of a typical binary system that leads to the formation of a BSS. The system is initially formed by a donor star of $1.02\ M_{\odot}$ and a companion of $0.81\ M_{\odot}$ with $P_{orb}= 0.80$~d. We assume that the evolution is conservative. The donor star, denoted with a black line, transfers mass to the companion when it is still burning its hydrogen core. The magenta line corresponds to the companion star. Finally, at an age of 14~Gyr, the donor and accretor have masses of $0.19065\ M_{\odot}$ and $1.64934\ M_{\odot}$ respectively, and an orbital period of $P_{orb}=1.66377$~d. Points labeled with A denote the beginning of the evolution, and B (C) stands for the onset (completion) of mass transfer. \label{Fig:BSS}}
\end{figure}

Here, we are especially interested in the binary scenario for forming BSSs. We should remark that to populate the BSS region of a given cluster, a binary should have a donor mass lower but closer to that corresponding to single stars on the turning point of the main sequence. Specifically, in Figure~\ref{Fig:BSS} we show the evolution of the components of a binary system initially formed by solar composition stars of $1.02\ M_{\odot}$ together with $0.81\ M_{\odot}$ companion, on a circular orbit of $P_{orb}= 0.80$~d. For simplicity, we assumed that the evolution is conservative and neglected the effects of rotation on stellar structure. Starting from point A, the donor star evolves faster, and at the age of 10.7~Gyr it fills its Roche lobe on point B of the black line. At this moment, the companion has evolved slightly (point B on the magenta line) and, due to accretion, it begins to get brighter. Sometime later, the companion becomes the brighter component of the system. When the donor detaches from its Roche lobe, at point C with an age of 12.5~Gyr, it evolves bluewards, becoming a pre-white dwarf with $0.19065\ M_{\odot}$. At C, the companion has ended its accretion with a mass of $1.64934\ M_{\odot}$. The final portion of its track resembles that of an isolated star. Evolution goes on up to the age of 14~Gyr. The final orbital period is of $P_{orb}=1.66377$~d.

As BSSs are far from us, we observe a single, unresolved object. So, the evolutionary track relevant for the observer is given, for a photometric system, by the addition of the light of both stars. For such a purpose, it is necessary to compute model atmospheres for each evolutionary stage. This procedure has been performed by \citet{Rain_BSSs} for a suite of open clusters. Notice that Figure~\ref{Fig:Messier3} is color~vs.~apparent magnitude in the V filter, while Figure~\ref{Fig:BSS} is a temperature~vs.~luminosity plane. For the transformation of the theoretical results to magnitudes and color index, see Figure~4 of \cite{Rain_BSSs}.

No doubt, the binary scenario for forming BSSs is promising but still suffers some serious drawbacks. In the simulation presented in Figure~\ref{Fig:BSS} we assumed conservative mass transfer ($\beta=1$); i.e., all mass lost by the donor is accreted by the companion, becoming the BSS. Of course, if some matter were lost by the system ($\beta<1$), the companion would be dimmer. Again, in this context, $\beta$ is a free parameter. So, we cannot claim we can predict quantitatively the system's behavior and what would be observed.

In Figure~\ref{Fig:BSS}, we have not considered the effects of rotation on stellar structure. This approximation should be relaxed for the next generation of models. When the companion accretes mass, it also gains angular momentum, speeding up its rotation. {\it Then, BSSs due to mass transfer in binary systems should be fast rotators}. Rotation leads to the occurrence of meridional circulation that modifies the chemical profile of the stars \citep{Zahn92}. Moreover, an accretion disk should be present around the accreting companion. But here, another fundamental question has to be stated. What is the fraction of mass retained by the companion if it rotates near critically? If we assume that the star and the surrounding accretion disk are polytropic\footnote{I.e., that the relationship between pressure and density is given by a power law $P \propto \rho^{1+1/n}$ where $P$ is the pressure, $\rho$ is the density and $n$ corresponds to the polytropic index}, a star can gain mass even if rotates critically \citep{Paczynski_Disco}. However, real stars are not polytropic.

Another important question is related to systems that reach a contact configuration. What is their subsequent evolution? The resulting object has to be observable among the components of the cluster to which it belongs. This remains an open question.

\section{Conclusions}\label{Sec:conclu}

In this paper, we have presented and discussed some general characteristics of the evolution of stars belonging to close binary systems (CBSs) that undergo mass transfer by Roche lobe overflow. We have paid particular attention to synthetize and expose the work developed in our research group {\tt GESBI}. In particular, we have centred our attention on two very different kind of systems. One is the case of systems that contain a compact object (a NS or a BH) and behave as X-ray sources, in particular those of low mass, usually called Low Mass X-ray Binaries (LMXBs) and the group of spiders related to it. The other addressed case is that of Blue Straggler Stars (BSSs) whose population is currently interpreted as due, at least partially, to binary evolution.

Stellar evolution in CBSs provides a natural reference frame for understanding the occurrence of all the objects we have discussed here. This should be considered as a success of the theory. However, it is true that in all models, we have had to consider some phenomena with a parametric description among the ingredients. Even for the case of the simplest systems, LMXBs without IFB or evaporation, we have included two parameters that describe the fraction of mass ($\beta$) lost by the donor star that the companion retains and the specific angular momentum of the mass lost from the system ($\alpha$). To be in conditions to construct more physically plausible models, it is of key importance to develop a theory capable of treating this phenomenon in a parameter-free way. This type of consideration applies to all the topics presented in this paper.

In any case, it is very interesting that many binary pulsars cannot be simply interpreted in the frame of non-irradiated models or without evaporation. These physical effects tie up the group of spiders, unifying the description and giving reasons to understand the known systems (including Redbacks, Black Widows, and Huntsman; \citealt{Rambo}). Also, it is remarkable that at least in the case of V404~Cyg, an initially non-rotating BH cannot be spun up only by mass transfer. We are currently studying if this is also the case for other LMXBs containing BHs.

In the case of BSSs, the evolution in CBS provides a natural scenario to make the companion, accreting star become the most massive and luminous of the pair during or after the mass transfer episode. In this case, the parameter $\beta$ related to how much mass is retained by the companion is of key relevance. It largely determines the luminosity the accreting star can reach. There is strong observational evidence for the existence of two BSS sequences in clusters, for example, in the globular cluster Messier~30 \citep{BSS_dossecuencias}. Sequences are usually interpreted as one due to mass transfer and the other due to stellar collisions. Collisions are probable in globular clusters, but these sequences are also present in open ones in which the stellar density is far lower, and consequently, collisions should be rare. But, is it impossible to accommodate both sequences as due to binary evolution in a single plausible description? It remains as an open question.

\section*{Acknowledgments}

The authors want to acknowledge IWARA Organizing Committee for the invitation to write this article. OGB wants to acknowledge Giovanni Carraro for calling his attention to the problem of BSSs, to Leonardo Benvenuto and Gabriel Ferrero for their assistance in preparing some figures. OGB is a member of the Carrera del Investigador Científico of the Comisión de Investigaciones Científicas of the Provincia de Buenos Aires (CICPBA), Argentina. MADV is a member of the Carrera del Investigador Científico of CONICET, Argentina. LBK and MLN are doctoral fellows of CONICET. JEH wishes to thank the GARDEL Group members at the IAG-USP for discussions and advice, and the Fapesp Agency (S\~ao Paulo State) and CNPq (Federal Government, Brazil) for financial support through grants and fellowships.  \\
Co-funded by the European Union (ERC-2022-AdG, {\it "StarDance: the non-canonical evolution of stars in clusters"}, Grant Agreement 101093572, PI: E. Pancino). Views and opinions expressed are, however those of the author(s) only and do not necessarily reflect those of the European Union or the European Research Council. Neither the European Union nor the granting authority can be held responsible for them.



\subsection*{Financial disclosure}

None reported.

\subsection*{Conflict of interest}

The authors declare no potential conflict of interest.





\nocite{*}
\bibliography{GESBI}%



\end{document}